\documentclass[twocolumn,showpacs,amsmath,amssymb,prl,superscriptaddress,floatfix]{revtex4}
\usepackage{graphicx}
\usepackage{bm}
\usepackage{amsmath,amsfonts}

\begin{document}

\title{Thermalization in a coherently driven ensemble of two-level systems}

\author{Igor Lesanovsky}

\author{Beatriz Olmos}

\author{Juan P. Garrahan}

\affiliation{School of Physics and Astronomy, University of
Nottingham, Nottingham, NG7 2RD, UK}

\date{\today}

\begin{abstract}
We study the coherent quantum evolution of a closed and driven mesoscopic chain of two-level systems that interact via the van-der-Waals interaction in their excited state. The Hamiltonian consists of a part corresponding to a classical lattice gas and an off-diagonal driving term without classical counterpart. We show that in a certain parameter range the latter leads to a thermalization of the system with respect to observables of the classical lattice gas such as the interaction energy and particle number distribution. We investigate the evolution of the system into this thermal state and discuss how to determine the corresponding temperature. Our findings can be applied to understand thermalization in strongly interacting systems of laser-driven Rydberg atoms, ions or polar molecules.
\end{abstract}
\maketitle

One of the most intriguing questions in many-body physics is how and under what conditions a closed quantum system evolves into thermal equilibrium. This problem has recently received much attention and the approach of a closed system to equilibrium (or the absence of it) has been subject of a number of experimental and theoretical efforts \cite{Kinoshita06, Hofferberth07,Rigol07,Rigol08,Biroli09,Olmos10,Goldstein10,pal10}. The former have been predominantly driven by the progress in ultracold atomic physics which allow to study coherent many-body dynamics in a very clean and decoherence-free environment \cite{Bloch08}.

In this work we investigate the thermalization of a mesoscopic ensemble of strongly and long-range interacting (laser-)driven two-level systems represented by atoms \cite{Weimer08,Sun08,Olmos10}, ions \cite{Muller08} or polar molecules \cite{Micheli06} confined in a deep lattice. The external dynamics is frozen out and the Hamiltonian of the internal dynamics can be decomposed into a classical part $H_\mathrm{LG}$ corresponding to a lattice gas and a quantum part $H_\mathrm{Q}$ originating from the driving.  We show that in a certain parameter regime the system reaches a steady state after an initial transient period.  Here observables of the classical lattice gas, such as the mean number of excited particles and the interaction energy, have thermalized according to a canonical distribution with of temperature $T$. We show that $T$ is proportional to the strength of the laser driving, and outline two ways for its experimental determination. The first method relies on measuring the statistics of the interaction energies.  The second is based on determining the response of the number of excited particles to a change of the chemical potential. Both methods are applicable to recent experiments with ultracold and highly excited atoms \cite{CubelLiebisch05,Heidemann07,Reetz-Lamour08}.

We aim for our analysis to stay close to experimentally realizable conditions.   We therefore consider
initial conditions which can be achieved and observables which can be accessed in actual experiments. We hope that our results here will motivate new experimental research to shed light onto the general issue of thermalization.

Our system consists of a periodic chain of $L$ driven two-level systems which - once excited - interact via the van-der-Waals interaction which decays with the inverse sixth power of the interparticle distance. The Hamiltonian is given by $H=H_\mathrm{Q}+H_\mathrm{LG}-\mu\sum_{k=1}^L n_k$ with
\begin{align}
  H_\mathrm{Q}=\Omega\sum_{k=1}^L \sigma^{(k)}_x&,&H_\mathrm{LG}= \frac{V}{2} \sum_{k\neq m}^L \frac{n_m n_k}{|m-k|^6}.\label{eq:Hamiltonian}
\end{align}
The ground state $\left|g\right>_k$ of each particle is coupled to its excited state $\left|e\right>_k$ via the operator $\sigma^{(k)}_x=\left|e\right>_k\!\left<g\right|+\left|g\right>_k\!\left<e\right|$, at a rate (Rabi frequency) $\Omega$. This form is typical for a laser-atom interaction treated within the rotating wave approximation. The operator $n_k=\left|e\right>_k\!\left<e\right|$ projects on the excited state of the $k$-th particle. The energy scale of the interaction is given by $V=C_6/a^6$ where $C_6$ is the van-der-Waals coefficient \cite{Singer05,Micheli07} and $a$ is the (regular) spacing between the two-level systems. The quantity $\mu$ corresponds to the chemical potential which in practice can be adjusted by varying the detuning of the laser that drives the transition: $\left|g\right>\rightarrow\left|e\right>$.

Throughout this paper we focus on the coherent evolution of the system from the initial (product) state $\left|\mathrm{init}\right>=\bigotimes_{k=1}^L\left|g\right>_k$, as this state is the natural starting point from an experimental perspective.  We are interested in the approach of the system to equilibrium. Since the system is closed, and evolves coherently, equilibration is not expected to occur with respect to any arbitrary observable \footnote{In fact, one can always construct a (very complex and non-local) observable that will not equilibrate.}.  Here we focus on observables of the lattice gas represented through $H_\mathrm{LG}$.  Such 'classical' observables are the total number of excited particles, $N=\sum_k\,n_k$, or the density-density correlations, $g_{nm}=L^2\left<n_m n_n\right>/\left<N\right>^2$.

\begin{figure}
\includegraphics[width=9.cm]{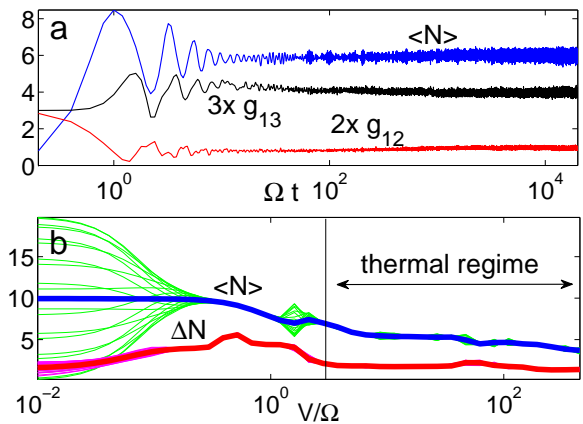}
\caption{(a) Temporal evolution of the mean excitation number $\left<N\right>(t)$ and the correlation functions $g_{12}(t)$ and $g_{13}(t)$ ($L=20$, $V=5\Omega$, $\mu=0$).  All quantities reach their steady-state - characterized by small amplitude oscillations - in the interval $10^1 < \Omega t < 10^2$. (b) Expectation value of the number of excited particles $\left<N\right>$ (blue) and its fluctuations $\Delta N$ (red), studied over the time interval $47<\Omega t<50$, as a function of the interaction strength. The thin lines correspond to $20$ instantaneous snapshots taken during that period. Thick solid lines represent the average over that time interval. For very small $V$ no relaxation into a steady state is occurring as large oscillations of the $\left<N\right>$ and $\Delta N$ are present. For large values of $V$ the instantaneous values of $\left<N\right>$ and $\Delta N$ differ only very little from the averaged ones. Moreover, the fluctuations $\Delta N$ are small. We refer to this as the thermal regime.} \label{fig:equ_quantitities}
\end{figure}

In Fig.\ \ref{fig:equ_quantitities}(a) we show the time-evolution of these lattice gas quantities for the parameter choice $V=5\,\Omega$ for which equilibration seems to occur. The observed behavior is the same for all lattice gas observables shown:  The initial large amplitude oscillations diminish in the interval $10^1 < \Omega t < 10^2$,  and for even longer times they are drastically reduced---the subsystem probed by these observables appears to have reached a steady state. Similar behavior has been reported for other systems such as hard core bosons confined to a lattice \cite{Rigol09}. In the following we will mainly focus on the expectation value of $N$ with respect to which the system reaches a steady state at $\Omega\, t \approx 50$ for the parameters of Fig.\ \ref{fig:equ_quantitities}(a). All times considered in this work are smaller than the quantum mechanical revival time.

Our aim is now to analyze why and under which circumstances $\left<N\right>$ equilibrates.  We first note that $\left<N\right>$ reaches a steady state value only if the interaction energy is sufficiently strong.
To show this we propagate the initial state up to a final time $\Omega\,t_\mathrm{f}=50$ and monitor the final value of the expectation of the number of excited atoms, $\left<N\right>$, and its variance, $\Delta N= \sqrt{\left<N^2\right>-\left<N\right>^2}$, for different interaction strengths $V$.   The corresponding plot is presented in Fig.\ \ref{fig:equ_quantitities}(b). The thick curves result from averaging over the time window $47<\Omega\,t<50$ at any given value of the interaction strength. Thin lines show instantaneous snapshots taken during this time interval. For $V=0$ we find large amplitude oscillations of the snapshots which differ strongly from the mean value $\left<N\right>\approx 0.5\,L$. In this non-interacting regime no relaxation into a steady state occurs for the lattice gas subsystem.  For increasing $V$, but still with $V\ll \Omega$, the instantaneous fluctuations decrease compared to the mean. In this regime the dynamics is approximately governed by an integrable Hamiltonian as shown in Ref. \cite{Olmos09-3}. Although $\left<N\right>$ becomes stationary we do not consider this to be a thermal state since its fluctuations $\Delta N$ are large. Moreover, we will later see that a temperature cannot be meaningful defined in this weakly interacting regime.

More interesting for us is the strongly interacting regime. It is located in Fig.\ \ref{fig:equ_quantitities}(b) right after the drop in $\Delta N$ which occurs at $V\approx 2\Omega$.  In this regime Hamiltonian (\ref{eq:Hamiltonian}) is no longer analytically solvable.  Here the driving term $H_\mathrm{Q}$ leads to the thermalization of observables related to the lattice gas.  In particular, we show below that the probability $p_n$ to measure $n$ excited particles is given by $p_n=\sum_\varepsilon\,p_{n\varepsilon}\propto\sum_\varepsilon \exp{S(n,\varepsilon)}\, \exp{-\beta\varepsilon}$.  Here $\beta$ is an inverse temperature, $\beta \equiv 1/ k_{\rm B} T$ (in what follows we set the Boltzmann constant to unity, $k_B = 1$), and $S(n,\varepsilon)$ is an entropy function obtained from the number of states with $n$ particles located in a given interval of the interaction energy $\varepsilon$.

\begin{figure}
\includegraphics[width=8.5cm]{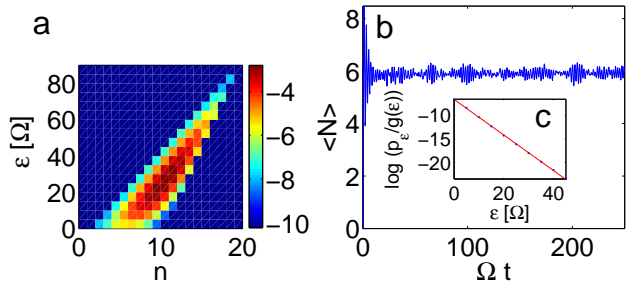}
\caption{(a) Entropy function for $L=20$. Shown is $S(n,\varepsilon)-\log\mathrm{dim}$, where $\mathrm{dim}$ is the total number of states of the system. The interaction energy is coarse grained over the interval $\varepsilon_k=[k\,V-V/2,\,k\,V+V/2]$. (b) Evolution of $\left<N\right>$ into equilibrium for $V=5\,\Omega$.  (c) The logarithm of $p_\varepsilon/g(\varepsilon)$ is given by the dotted line. The red solid line represents a linear fit.  } \label{fig:equilibrium}
\end{figure}

To prove this we proceed as follows.  We define the projector $P_{n\varepsilon}=P_n \otimes P_\varepsilon$, where $P_n$ projects onto the subspace containing $n$ particles, and $P_\varepsilon$ projects onto the subspace of states which possesses the interaction energy $\varepsilon$. As in the numerics we have to deal with a finite system, and thus a non-continuous density of states, we define a coarse-grained interaction energy: We introduce energy intervals that are centered around $\varepsilon_k=k\,V$ (with $k=0,1,...,L$) and have width $V$. Each of these energy intervals contains a large number of (quasi) degenerate states onto which  $P_\varepsilon$ projects.  We now transform into the interaction picture with the transformation $U=\exp\left(-i\,H_\mathrm{LG}\,t\right)$, where the Hamiltonian for $\mu=0$ is given by $H_x(t)=U^\dagger H_\mathrm{Q} U= \sum_{\varepsilon\varepsilon^\prime} P_\varepsilon H_\mathrm{Q} P_{\varepsilon^\prime} \exp(-i\omega_{\varepsilon\varepsilon^\prime}t)$ with $\omega_{\varepsilon\varepsilon^\prime}=\varepsilon^\prime-\varepsilon$.
The evolution of the density matrix is then given by the von-Neumann equation $\partial_t\rho=-i\left[H_x(t),\rho\right]$ which is equivalent to
\begin{eqnarray*}
  \frac{\partial\rho(t)}{\partial t}=-i\left[H_x(t),\rho(0)\right]-\int_0^t\!\! ds \left[H_x(t),\left[H_x(s),\rho(s)\right]\right].
\end{eqnarray*}
We can now to derive a closed equation for the evolution of $p_{n\varepsilon}=\mathrm{Tr}\, P_{n\varepsilon} \rho(t)$. By generalizing the ideas from Ref.\ \cite{Olmos10} we find that this equation is given by
\begin{eqnarray*}
  \frac{\partial p_{n\varepsilon}}{\partial t}&=&\sum_{\varepsilon^\prime m}\int_0^t d\tau\,\cos[\omega_{\varepsilon\varepsilon^\prime}(t-\tau)]\times\label{eq:dP_t}\\
  && \left[\kappa_{nm}(\varepsilon,\varepsilon^\prime)\,p_{m\varepsilon^\prime}(\tau)-\kappa_{mn}(\varepsilon^\prime,\varepsilon)\,p_{n\varepsilon}(\tau)\right]\nonumber.
\end{eqnarray*}
Here we have exploited the fact that due to the strong interaction among the particles the couplings between subspaces with different particle number $n\neq m$ and interaction energy $\varepsilon\neq\varepsilon^\prime$ are uncorrelated.  The mean strength of these couplings is contained in the coefficients $\kappa_{nm}(\varepsilon,\varepsilon^\prime)$. We are now interested in coarse-graining this evolution equation over time intervals that are larger than $\Omega^{-1}$.  After this procedure one obtains
\begin{eqnarray*}
  \frac{\partial p_{n\varepsilon}}{\partial t}&\propto&\sum_{m}u_{mn}(\varepsilon)\left[\mathrm{dim}_{n}(\varepsilon)p_{m\varepsilon}-\mathrm{dim}_{m}(\varepsilon)p_{n\varepsilon}\right],
\end{eqnarray*}
where the coefficients $u_{mn}(\varepsilon)$ contain the coupling rates between subspaces with fixed interaction energy but different particle number. The quantity $\mathrm{dim}_{n}(\varepsilon)$ is the number of states that lie in the interaction energy window centered around $\varepsilon$ and contain $n$ excited particles. Provided that $u_{mn}(\varepsilon)\neq 0$ the steady state solution of this equation is of the form
\begin{eqnarray*}
  p_{n\varepsilon}=b_\varepsilon\,\mathrm{dim}_{n}(\varepsilon)=b_\varepsilon\,\exp{S(n,\varepsilon)}.
\end{eqnarray*}
The entropy $S(n,\varepsilon)$ is shown in Fig.\ \ref{fig:equilibrium}(a).
Since subspaces with different interaction energy are independent and uncoupled the probability to have $n$ excitations is given by $p_n=\sum_\varepsilon p_{n\varepsilon}=\sum_\varepsilon b_\varepsilon \,\exp{S(n,\varepsilon)}$.

What remains is to determine the functional dependence of the coefficients $b_\varepsilon$ on the interaction energy from numerical simulations.  As an illustration we consider a system with $L=20$ and $V=5\Omega$ whose temporal evolution of the mean particle number is shown in Fig.\ \ref{fig:equilibrium}(b). We study the system at $t=250\,\Omega^{-1}$ where the steady state is well-established. In order to determine the $b_\varepsilon$'s we consider $p_\varepsilon=\sum_n p_{n\varepsilon}=b_\varepsilon\,\sum_n \exp{S(n,\varepsilon)}=b_\varepsilon\, g(\varepsilon)$
where $g(\varepsilon)$ is the number of states with energy $\varepsilon$, and study the ratio $p_\varepsilon/g(\varepsilon)$. In the Fig.\ \ref{fig:equilibrium}c the logarithm of this quantity is plotted as a function of $\varepsilon$ which is excellently fitted by a Boltzmann distribution, and we find $ b_\varepsilon\propto \exp -\beta\varepsilon$ with $\beta=0.376\,\Omega^{-1}$.

\begin{figure}
\includegraphics[width=8.5cm]{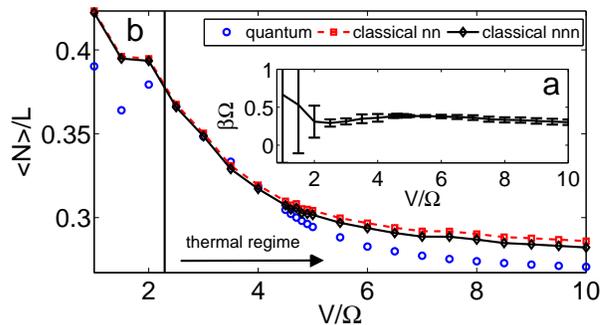}
\caption{(a) Inverse temperature as a function of the interaction strength after a propagation time of $\Omega t=250$ ($L=20$). The error bars delimit the $95\%$ confidence interval. (b) Mean density $\left<N\right>/L$ of the classical lattice gas in thermal equilibrium plotted as a function of the interaction strength $V$ (red/dash curve). The inverse temperature was taken from panel (a). The black solid curve shows the result if next-nearest neighbor interactions are accounted for perturbatively. For comparison the mean density which is obtained from the quantum calculation is shown (blue circles).} \label{fig:class_vs_quantum}
\end{figure}

The numerical results strongly suggest that at $V=5\,\Omega$ the steady state of the system is given by a thermal (canonical) distribution of the lattice gas Hamiltonian $H_\mathrm{LG}$ with an inverse temperature $\beta$ that is determined by the Rabi frequency $\Omega$.  To confirm if this also holds for other interaction strengths $V$ we determine the temperature as a function of $V$ according to the procedure outlined above. The corresponding data is shown in Fig.\ \ref{fig:class_vs_quantum}(a).  Below $V=2\Omega$
we see that the Boltzmann distribution gives a poor account of the numerical data, i.e., the fit of $b_\varepsilon$ to a function of the form $\exp-\beta\epsilon$ results in huge errors. For higher values of $V$ the fit works, which is consistent with the observations of Fig.\ \ref{fig:equ_quantitities}(b).

We now consider whether the above temperature measure gives results that are consistent with the expectation values obtained in a statistical mechanics way.  We do so by calculating the excitation density in the thermal ensemble of the lattice gas at the estimated temperature and comparing this to the actual excitation density from the quantum problem.  The partition function of the lattice gas is given as $\Xi(\beta,\mu,L)=\sum_{\{n_i\}} \exp -\beta \left[H_\mathrm{LG}-\mu\sum_{k=1}^L n_k\right]$.  The thermal average of an observable $O$ is given by $\left<O\right>_\mathrm{th}=\Xi^{-1}(\beta,\mu,L)\,\sum_{\{n_i\}}O\,  \exp -\beta \left[H_\mathrm{LG}-\mu\sum_{k=1}^L n_k\right]$. In the case of nearest-neighbor interactions on a ring, $\Xi(\beta,\mu,L)$ can be calculated straightforwardly via the transfer-matrix method \cite{Goldenfield92}, and we use this result neglecting contributions from interactions over a longer range. When necessary these interactions can be included perturbatively, as they are smaller by at least a factor of $1/64$ than the dominant nearest-neighbor contribution.  The partition function is given by  $ \Xi(\beta,\mu,L)=\lambda_+^L+\lambda_-^L$
with $\lambda_\pm \equiv \frac{1}{2} \left( 1 +e^{\beta(\mu-V)} \right) \pm \frac{1}{2} \sqrt{4 e^{\beta\mu}+\left(1-e^{\beta(\mu-V)}\right)^2}$.  In the limit $L\rightarrow\infty$, which is a good approximation for $L=20$, the excitation density at $\mu=0$ (i.e., excitation laser on resonance) obtained from this partition sum takes the simple form $\left<N\right>_\mathrm{th}/L=\frac{1}{2} +\frac{e^{-\beta V}-1}{2 \sqrt{4+\left(e^{-\beta V}-1\right)^2}}$.

In Fig.\ \ref{fig:class_vs_quantum}(b) we compare the excitation density that is obtained form the quantum calculation to the equilibrium value in a classical lattice gas at inverse temperature $\beta$ taken from Fig.\ \ref{fig:class_vs_quantum}(a). The agreement is good, but systematic deviations are evident beyond $V\approx4\,\Omega$.  These deviations are reduced if next-nearest neighbor interactions are included perturbatively.  Other initial states different from the one considered here are expected to lead to a better agreement with the lattice gas prediction, in analogy with the observations of Ref. \cite{Olmos10}.

Our findings indicate that the temperature inferred from the distribution of the interaction energy is compatible with a classical thermal ensemble. This temperature can be estimated in an alternative way, which corroborates our claim that the system has reached a thermal state with respect to observables of the lattice gas. This is done by comparing the response of the system to a change of the chemical potential with its spontaneous fluctuations. In thermal equilibrium the fluctuation-dissipation theorem states that
the change in the average number of excitations is given by its variance, i.e.,
\begin{eqnarray}
  \frac{\partial \left<N\right>_\mathrm{th}}{\partial \mu}=\beta \left[\left<N^2\right>_\mathrm{th}-\left<N\right>_\mathrm{th}^2\right],\label{eq:fluctuation_dissipation}
\end{eqnarray}
the constant of proportionality being the inverse temperature.   Experimentally such fluctuations have been measured for example in frozen Rydberg gases \cite{CubelLiebisch05} where they are used to characterize the relative importance of interactions \cite{Ates06}.  We are thus confident that the method we outline below will turn out to be a valuable tool for the experimental determination of the temperature.

\begin{figure}
\includegraphics[width=8cm]{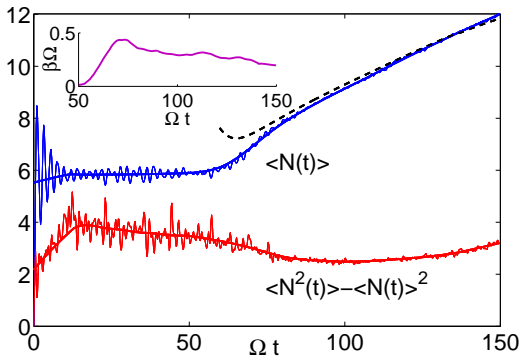}
\caption{Mean number of excited spins, $\left<N\right>$ (blue), and its variance, $\left<N^2\right>-\left<N\right>^2$ (red) for $L=20$. At $\Omega t=50$ the chemical potential $\mu$ is ramped up linearly from $0$ to $8\Omega$.  The thick curves result from smoothing the data. The inset shows the inverse temperature calculated from the fluctuation-dissipation theorem (\ref{eq:fluctuation_dissipation}).  The dashed curve is calculated using the classical lattice gas model at a chemical potential $\mu(t)$ and the corresponding temperature of the inset.} \label{fig:fluctuations}
\end{figure}

In Fig.\ \ref{fig:fluctuations} we show a numerical simulation of the envisaged experiment.
We start the Hamiltonian evolution of the system with $\mu=0$, keeping it that way until $\Omega t=50$ for the system to equilibrate.  We then linearly ramp up the chemical potential to $\mu_\mathrm{max}=8\Omega$ during a time interval of length $100\,\Omega^{-1}$.  Experimentally this is done by decreasing the detuning of the excitation laser which, as expected, leads to an increase of the mean number of excited particles. The parameters are chosen such that no saturation of the excitation number occurs.  We determine the fluctuations $(\Delta N)_\mathrm{th}^2=\left<N^2\right>_\mathrm{th}-\left<N\right>_\mathrm{th}^2$, and obtain the inverse temperature from Eq.\ (\ref{eq:fluctuation_dissipation}), using that $\partial_\mu \langle N(t) \rangle_{\rm th} = [\frac{d \mu(t)}{dt}]^{-1} \partial_t \langle N(t) \rangle_{\rm th}$.  The (time-dependent) temperature is plotted in the inset of Fig.\ \ref{fig:fluctuations}. At the beginning of the ramp ($\Omega t=50$) the density does not respond immediately to the change of $\mu$.  This finite time effect causes a spurious initial increase of $\beta$ from zero which saturates at $\Omega t \approx 70$ at $\beta=0.44\,\Omega^{-1}$. For consistency, this value has to be close to the value predicted from the distribution of the interaction energies at the beginning of the ramp, which is $\beta=0.40\,\Omega^{-1}$. Given the strongly fluctuating data and the data smoothing the agreement is good.
The inset of Fig.\ \ref{fig:fluctuations} also shows that with increasing $\Omega t$, and hence increasing $\mu$, the inverse temperature $\beta$ decreases. The system is heated, showing that the evolution is not fully adiabatic. For comparison we display $\left<N(t)\right>_\mathrm{th}$ calculated from the classical partition function with $\mu(t)$ and $\beta(t)$ (dashed line). The agreement is excellent.

The fact that both schemes suggested here give consistent estimates for the temperature is non-trivial. It indicates that the subsystem of the closed quantum problem that corresponds to the lattice gas observables is indeed in thermal equilibrium. This shows that aspects of the real time dynamics (for long times) of a strongly interacting and coherently driven quantum many-particle system can be understood through classical equilibrium thermodynamics.

\end{document}